\documentclass[10pt,conference]{IEEEtran}
\usepackage{wrapfig}

\usepackage{algorithm2e} 
\usepackage{amsmath}
\usepackage[american]{babel}
\usepackage{booktabs}
\usepackage{caption}
\usepackage{color}
\usepackage{comment}
\usepackage[keeplastbox]{flushend}
\usepackage[acronym]{glossaries}
\usepackage{listings}
\usepackage{multirow}
\usepackage{rotating}
\usepackage{siunitx}
\usepackage{soul}
\usepackage{tabularx}
\usepackage{tikz}
\usepackage[textsize=scriptsize]{todonotes}
\usepackage{url}
\usepackage{xspace}
\usepackage{graphicx}
\usepackage[export]{adjustbox}
\usepackage{subfigure}
\usepackage{enumitem}
\usepackage{authblk}
\newcommand{\toolname}{\textit{Android-COCO}\xspace}
\lstdefinelanguage
[x64]{Assembler}     % add a "x64" dialect of Assembler
[x86masm]{Assembler} % based on the "x86masm" dialect
{morekeywords={CDQE,CQO,CMPSQ,CMPXCHG16B,JRCXZ,LODSQ,MOVSXD, %
		POPFQ,PUSHFQ,SCASQ,STOSQ,IRETQ,RDTSCP,SWAPGS, %
		rax,rdx,rcx,rbx,rsi,rdi,rsp,rbp, %
		r8,r8d,r8w,r8b,r9,r9d,r9w,r9b, %
		r10,r10d,r10w,r10b,r11,r11d,r11w,r11b, %
		r12,r12d,r12w,r12b,r13,r13d,r13w,r13b, %
		r14,r14d,r14w,r14b,r15,r15d,r15w,r15b}} % etc.

\title{\toolname: Android Malware Detection with Graph Neural Network for Byte- and Native-Code}

\author[1]{Peng Xu}
\affil[1]{Technical University of Munich}

\begin{document}
	\maketitle

\begin{abstract}With the popularity of Android growing exponentially, the amount of malware has significantly exploded. It is arguably one of the most viral concerns on mobile platforms. Diverse methods have been introduced to detect Android malware based on \textit{Manifest} features (e.g., permissions, intent-filters, etc.). Unfortunately, it is infeasible for these works to detect malware applications that obfuscate their malicious byte- and native-code instructions. Existing research on malware has revealed that the calling associations between routines in a program can deliver in-depth data from which to develop a reliable malware detector. 

As a solution, we propose Android-COCO, a supervised approach for detecting Android malware. We automatically transform the byte and native code to program dependency graphs (PDG) and use the embedding of these graphs to classify applications. Large-scale experiments on 100,113 samples (35,113 malware and 65,000 benign) show that the our approach can detect malware applications with an accuracy of 99.86\%, significantly outperforming state-of-the-art solutions.
\end{abstract}

%%
%% The code below is generated by the tool at http://dl.acm.org/ccs.cfm.
%
% Please copy and paste the code instead of the example below.
%%
%%
%% Keywords. The author(s) should pick words that accurately describe
%% the work being presented. Separate the keywords with commas.
%\keywords{Android Malware detection, graph neural networks, ensemble learning, instruction2vec, basicblock2vec, function2vec}

%\newpage
\section{Introduction}
The mobile phone has become a necessity in our daily life. Due to its versatility, smartphone usage has grown exponentially to reach various sectors like social networks, e-commerce and banking. According to statistics on mobile OS market shares~\cite{smartphonemarketshare1}, Android is the world’s dominant mobile operating system, leading the market with over 83\% market share in comparison to iOS and Windows. Android’s enormous dominance is largely a function of being open source as well as providing users with a rich store that involves millions of various applications. Further, the brisk prosperity of smartphones along with Android’s huge popularity has not gone disregarded by malware authors. Rather, Android has become an attractive platform for launching thousands of malicious applications. Recently, the number of malware has increased by leaps and bounds to reach a disturbing level. Based on McAfee's report~\cite{mcafee}, the number of discovered Android malware has touched 3.5 millions in 2021, and each more there are around 300,000 new Android malware,  which has led the overall mobile malware's tally to reach 25 millions. 

To defend against the threat and mitigate the alarming spread of malware, several anti-virus scanners have been provided by computer security companies such as Symantec and McAfee. Many of these products use either signature-based~\cite{chen2018tinydroid} or code matching~\cite{jang2016detecting} methods. Unfortunately, these techniques have proved their incapability to keep pace with the sophistication of malware.  Hence, Android malware detection has been, and still is of the essence for security researchers in both industry and academia.

In recent years, various research works~\cite{wei2014amandroid,arzt2014flowdroid,gordon2015information,li2015iccta,lu2012chex,anderson2011graph,xu2018cdgdroid,narayanan2016contextual,arp2014drebin,xu2021detecting,xufalcon,xu2021hawkeye,hybroid} 
have been proposed to detect malware based on machine learning and deep learning techniques. The majority of these approaches use either Android application's attributes collected from the Manifest File, specifically, Permissions, Intent-filters, and Providers. Unfortunately, these approaches can be easily bypassed by certain obfuscation techniques, such as Identifier Renaming, Code Hiding, Reflection, Dead-Code Insertion or Instruction Substitution~\cite{suarez2017droidsieve}\footnote{~\cite{suarez2017droidsieve} considers native code as a type of obfuscation. However, we separate it from general obfuscation techniques in Android and we mention it in the following}. All these obfuscation methods aim to alter the application's appearance but preserve its behavior, either by encrypting the values of packages and classes or embedding useless instructions to the program~\cite{dong2018understanding}. In a nutshell, while these Manifest-based detection approaches offer a supplementary security control to Android, they demonstrate failure against obfuscated applications. Accordingly, some approaches go one step further by considering the source code. Researchers have figured out that the high-level properties of the code, in particular, the code structure, such as the Function Call Graph (FCG) and API calls robust against obfuscation~\cite{hu2009large,suarez2017droidsieve}.

Although some of aforementioned works, like Amandroid~\cite{wei2014amandroid}, flowdroid~\cite{arzt2014flowdroid}, %DroidSafe~\cite{gordon2015information}, % IccTA~\cite{li2015iccta},CHEX~\cite{lu2012chex}, 
yield high accuracy results, they operate solely on the DEX bytecode, which does not lead to a thorough analysis of Android applications. A recent study shows that 86\% of Android applications involve native code, allowing this latter to be a reasonable threat vector~\cite{sun2014nativeguard}. That is to say, an application with a malicious embedded native code will go undetected by a layer that merely focuses on the DEX bytecode. This motivates the need for a byte-code and native-code combination approach that characterize and generalize the malicious patterns embedded in Android apps.
 
In this paper, we design an ensemble algorithm to combine the byte-code and native-code malware detection sub-systems(or layers). In particular, we present a layered approach that blends the structural and inter-language analysis of Android applications. %The aim of our method is to mitigate the possible gap of one layer by the strength of the other layer, allowing the framework to tackle every attack vector associated with Android applications. 
To this end, our approach consists of two layers, the graph-based detection layer and the ensemble layer.
For the graph-based detection layer, it includes two sub-systems:
\begin{itemize}
	\item Byte-code sub-system: It is based on the DEX byte-code. It first extracts the Program Dependence Graph (PDG) of the Android application, and then employs an efficient graph embedding to transform the graph's main features to vectors, which is used as an input for our neural network. In an evaluation with 100,113 samples (35,113 malware and 65,000 benign), our structural-based approach is shown to be immensely efficient, as it yields an accuracy of 99.8\%, which outperforms numerous existing works.
	\item Native-code sub-system: It is based on the native code (.so files). It employs the graph embedding technique to transform the dynamic libraries (.so files) to vectors, which is then used to feed our network-based classifier. Large scale experiments on the used datasets result in an accuracy rate of 96.66\%.
\end{itemize}
After the detection sub-systems, we design an ensemble algorithm as our second-layer, which combines the output of the two MLP classifiers.
\begin{itemize}
	\item Using the ensemble algorithm, we improve the accuracy to 99.86\%.
\end{itemize}
In summary, our primary contributions are described as follows:
\begin{itemize}
	\item \textbf{we introduce the graph embedding to Android malware detection.} We present a novel approach that employs graph embedding technology to differentiate between benign and malicious applications. We encode either the Program Dependence Graph (PDG) and function call graph(FCG) into its own vector representations. The graph's embedding captures the graph topology and hence covers every layer's specific features. To the extent of our knowledge, this is the first work that consider both the byte-code and native-code to detect Android malware.
	\item  \textbf{we use ensemble algorithm to combine the results of Android malware detection} We present a multi-layer approach that couples byte-code and native-code analysis from two different application components. The complementary strengths among the analysis of byte-code and native code's structure creates a thorough detection system.
\end{itemize}

\section{Background}
In this section, first, we throw the lights on the architecture of Android applications and the existing android malware detection techniques. Then, we introduce the used Natural Language Processing concepts, namely the embedding techniques as well as graph neural network.  
 
\subsection{Android Package Kit (APK)}\label{sec:apk}

Android applications are deployed in the form of Application Package Kit (APK), which is a zipped file that involves the app's code in the DEX file format, native libraries, resources, assets and AndroidManifest.xml file. 
As for the DEX file, it involves the application's code. As mentioned earlier, Android uses Dalvik virtual machine to execute applications. Hence, it uses Dalvik bytecode instead of Java bytecode. At first, the Java source code is compiled to Java bytecode by a Java compiler, which is then compiled to Dalvik bytecode by the DEXcompiler. 
A DEX file contains classes definitions, each class involves one or more methods'(functions') definitions. Each method(function) comprises of one or more basic block, which involves one or multiple instructions. As for the instruction's format, it consists of a single opcode and multiple operands. In overall, Dalvik has 256 opcodes. Android applications make use of 222 opcodes as the others are reserved for future usage~\cite{dalvikfig}.

%\subsection{Android Obfuscation}
%Obfuscation is the act of obscuring the source code to hinder the understanding for humans with preserving the functionality of the program. For security researchers, it is considered to be a mixed-blessing technique. To be precise, for a legal software company, obfuscation is used to prevent competitors from unfairly building software based on identical source code. However, for an Android malware writer, obfuscation is used to conceal malicious code patterns in a program to easily bypass Android malware analysis and detection tools~\cite{dong2018understanding,suarez2017droidsieve}. Recently, several works ~\cite{hoffmann2016evaluating,maiorca2015stealth} present results of new obfuscation frameworks that aim to break the malware analyzing by the machine learning based detection systems.

\subsection{Program Dependence Graph}\label{sec:pdg}
A Program Dependence Graph(PDG) is a graph that explicitly considers both data and control dependencies. Data dependence graph(DDG) represents the program’s appropriate data flow whereas the control dependence  graph (CDG) illustrates the program’s control flow relationships. The combination of both dependencies covers all the program’s computational parts, which leads to an efficient use when it comes to programs' optimizations. As for the graph’s topology, Program Dependence Graph’s nodes are represented as statements or instructions, while the edges represent both the data values and control conditions necessary to the execution of the node’s operations~\cite{ferrante1987program}. The PDG of a program P is the union of a pair of graph: the DDG of P and the CDG of P. Note that the DDG and CDG are subgraphs of a PDG. The PDG is obtained by merging the DDG and the CDG~\cite{kazmierski2011system}.
 
\subsection{Word, Sentence and Document Embedding}
\label{w2v}
Adapting real life scenarios into machine learning models entails the mapping of textual inputs or scripts into numerical representation that is semantically expressive. Further, to ensure a good representation of a text input, the numerical representation should catch the linguistic sense of every word in the dataset. To be more specific, an informative numerical representation can enormously impact the performance of the model.
There exists several natural language processing techniques to handle text data, such as word embedding~\cite {datatowardsscience, mikolov2013distributed}, sentence embedding~\cite {datatowardssciencesentence}, and document embedding~\cite {mikolov2013distributed}. 
\begin{itemize}
\item {Word Embedding}: It is a numeric vector representation, obtained by embedding semantic meanings of words from a particular corpus. Word2Vec is a technique to build such an embedding based on neural network. Word2Vec can be achieved through two techniques which employ neural network. Namely, Skip-Gram and Common Bag Of Words (CBOW). Skip-Gram learns the prediction of a target word based on its neighbors, whereas CBOW uses the context to predict a target word~\cite {datatowardsscience,mikolov2013distributed}.
\item {Sentence Embedding}: It follows the same concept as word embedding. It represents a sentence with a numeric vector representation. A direct and frequently used approach for embedding sentences into a vector space is to aggregate the embedding vectors of the words forming a sentence~\cite {datatowardssciencesentence}.
\item {Document Embedding}: It is an extension to Word2Vec. Word2vec learns to represent a given word to its numerical vector representation whereas Doc2Vec aims at representing a document into its numerical vector space~\cite {mikolov2013distributed}.
\end{itemize}
\subsection{Graph Embedding}

Nowadays, a graph is considered to be a rich input for analysis works. It has been commonly applied in various fields. Further, feeding a machine learning model with graphs is very challenging. Here, the graph embedding comes in handy and acts as a technique that turns the graph's properties to a set of numeric vector representations to feed a particular model. The rich the graph embedding is, the good the results can be achieved. Thus, the embedding should catch all the graph's significant information, which involve its topology, vertex-to-vertex relationships, and vertices. 

Graph embedding can be categorized into the following two types~\cite{datatowardssciencegraph}.
The first one is node embedding, which is used to perform either classification or prediction tasks on vertex-level. There exist numerous works that deal with the node embedding area, such as Deepwalk~\cite{perozzi2014deepwalk}, Node2vec~\cite{grover2016node2vec}%, SDNE~\cite{wang2016structural} 
and Graph Convolutional Networks (GCN)~\cite{kipf2016semi}. Meanwhile, the other is graph embedding, which is suitable when it comes to performing classification or prediction on a graph level. There are several techniques that can be adopted to perform graph embedding, such as Structure2Vec~\cite{song2018structure2vec} and Graph2Vec~\cite{narayanan2017graph2vec}.
\section{Motivation}
As we mentioned above, there are many works in the Android Malware Detection field, however, nearly all of works are concentrating on the Android Byte-code, which includes the functional behavior writing in Java language. Currently, there are more and more Android application embeds native libraries (written in C and C++ language) in them and the trend to detect and categorize Android apps becomes significant. 

Therefore, on the one hand, we need a methodology which is not only able to detect Android malware byte-code level, but also native code level. 

On the other hand, since the platforms to support Android OS and its ecosystem are multiple, and many 

\section{System Design}
Our system follows the principle of Detection In-Depth~\cite{detectionindepth}, which is a security strategy that protects an asset by implementing multiple layers of detection to avoid a single point of failure. In our context, to defend against the Android malware threat, we placed two detection models (sub-systems) that vary in terms of focus. 

To be precise, our first layer operates on the application's structural information, including byte-code sub-system and native code sub-system, whereas the second layer concentrates on the ensemble algorithm to combine the two sub-systems of the first layer. The advantage of coupling two detection layers is to ensure an extensive examination of different application's components. Additionally, our multi-layer approach deals merely with applications that involve native code. Regarding our system's output, an application with native code is considered to be benign only if both sub-systems of the first layer prove its benignity. Otherwise, it is considered a malware. For applications without native code, the classification result is fully based on the byte-code sub-system's analysis.

\begin{figure*}[h]
	\centering
	\includegraphics[scale=0.35,trim=2 2 2 2]{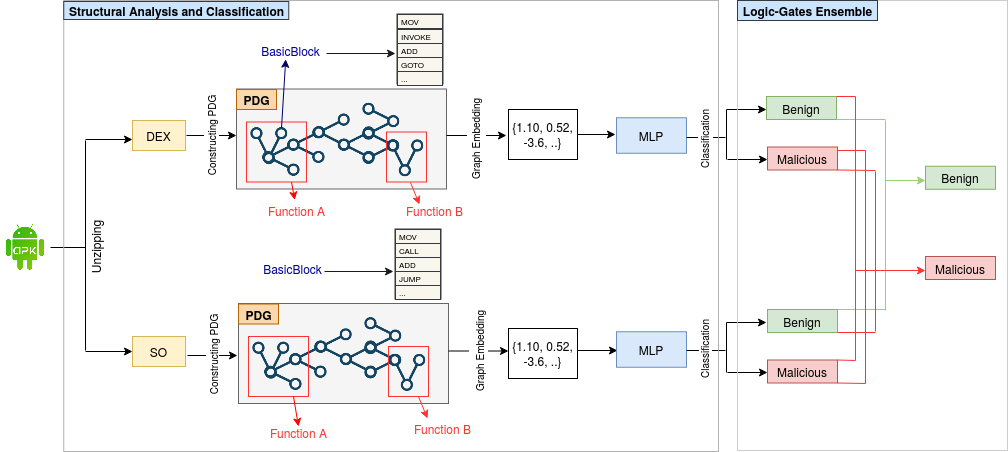}
	\centering
\caption{\toolname architecture}
	\label{struct}
\end{figure*}
We formalize our two-layer malware detection system as a binary classification problem. In the first layer, we define the graph sets, which represents the structural information of the executable files and native libraries (malware and benign), as our input, like \textbf{$G^D(V^D, E^D)$}, where $V^D$ presents the set of graph's nodes, and $E^D$ presents the edges among those nodes and D presents the number of graph. For each graph $g_i$($g_{ib}$ and $g_{in}$), it is encoded as $g_i(V,X,A)$, where $A \in \{0,1\}^{m * n}$ is the adjacency matrix and m is the number of the graph's node, $m = |V|$. $X \in R^{m * d}$ presents a d-dimensional real-valued vector $x_j \in R^d$. 
\begin{equation}
\begin{split}
{pred_{i}}_b = f_b(g_{ib}(V_{ib}, X_{ib}, A_{ib}) \\
{pred_{i}}_n = f_n(g_{in}(V_{ib}, X_{ib}, A_{ib}) \\ 
\min \sum_{i=1}^{D} loss_{ib} = \lambda({pred_i}_b , y_{ib}|c_{ib}) + \delta w(f_b)) \\ 
\min \sum_{i=1}^{D} loss_{in} = \lambda({pred_i}_n, y_{in}|c_{in}) + \delta w(f_n)) \\
%\min \sum_{i=1}^{D}  ({loss_i}_b + {loss_i}_n) 
\label{system_eq}
\end{split}
\end{equation}
The goal of the first layer of our system is to learn a mapping from the $G^D$ to $Y^D$, $f : G \to Y$ to predict whether an executable file or a native library is malicious or not. The prediction function f can be learned by minimizing the loss function in Eq.~\ref{system_eq}. For the second layer, we use the logic gates ensemble to combine the results of two classifiers in the first layer. We will implement for formalized model in the following, where $g_i$($g_{ib}$ and $g_{in}$) in Eq.~\ref{system_eq} is our graph embedding and $f$($f_b$ and $f_n$) in Eq.~\ref{system_eq} is our MLP-classifier(we use two MLP classifiers). 

\subsection{Byte-code based Malware Detection }
%\subsubsection{Byte-code malware detection}
The first subsystem of our approach deals with the DEX bytecode. %Usually, for optimum Android malware detection, a system demands to catch every aspect of an application. Besides the Manifest File, which involves only contextual information, an android application includes code, which may involve malicious parts. Consequently, a deep examination of the DEX files is necessary. Normally, malicious applications share comparable code structures. This is simply linked to sophisticated techniques used by malware authors to conceal malicious functionalities, either by embedding malicious code to benign applications using repackaging techniques~\cite{marastoni2017groupdroid}, or building malware applications from scratch in a way that does not reveal explicit malicious indicators. In this respect, obfuscation has appeared as a useful technique for hiding malicious payloads. Thus, malware authors have started to obfuscate their malicious content to easily circumvent existing malware detection. Unfortunately, malware's sophistication has lowered users' trust in benign applications, app stores, and possibly the whole Android ecosystem. 
%To this end, 
We designed our first subsystem to deal with the DEX bytecode of Android applications. Accordingly, to obtain the main features of the DEX bytecode, we need to get the execution paths that fully covers the program. Unfortunately, with traditional analysis, this could hardly be achieved. As a result, graph-based techniques have been introduced, as they produce rich data representations. Contrary to early graph-based malware detection approaches that employed control flow and function call graphs to detect obfuscated parts of the code, our approach utilizes the Program Dependence Graph (PDG), which applies both data and control analysis. To put it differently, a PDG combines an explicit representation of data and control flow relationships of an Android application, which exposes the full program structure. Although malware variants vary in terms of code appearance due to obfuscation, they share one common program structure, which can be plainly observed by analyzing their PDGs.

As illustrated in Figure~\ref{struct}, the structural byte-code subsystem is a five-step process. After the application's disassembling, we encoded every basic block to its corresponding vector and extracted the PDG. Next, we converted the entire PDG to a numerical vector of real numbers to feed our neural network.

\subsubsection{Opcode Embedding}

As described in Section~\ref{sec:apk}, Android applications are developed in Java and utilize a Dalvik virtual machine to execute. That is to say, an application's source code is first compiled to Java bytecode, then translated to Dalvik bytecode and saved in DEX. In the first subsystem of our malware detection approach, our aim is to tackle the DEX bytecode using the PDG, which models both control and data flow relationships. Consequently, after the disassembling phase, we merely considered Dalvik executable files. Further, capturing the PDG's topology requires handling both control and data dependencies found in an application. To this end, we treated every basic block of an application as a node. Additionally, we considered edges to express links bridging values' definitions and their uses to ensure a broad coverage of all the application's methods. Before casting the lights on the opcode embedding phase, it is important to understand the format of a basic block, which is a series of sequential instructions that are executed consecutively with no jumps in the middle~\cite{basicblock}. A basic block can involve one or more instructions. Each instruction is composed of two elements, opcode, and operand. Our aim is to capture the context of every instruction in a basic block. Thus, we embedded each instruction to its corresponding vector. 

%As demonstrated in Listing~\ref{codetransformation}, the above Java code is transformed into 4 basic blocks. Each basic block contains a series of statements that are executed sequentially. Next, those statements are first converted to Java bytecodes then to Dalvik bytecodes. As for the embedding process, given that opcode makes the first byte of instruction and specifies the operation to be carried out by the hardware, 
We used opcode as an our input to build our corpus of vectors using the Word Embedding technique. Regarding the implementation of this step, we first built a large unsupervised corpus of opcodes extracted from our datasets. To this end, we used Androguard to retrieve an application's set of opcodes. Androguard has a function AnalyzeAPK that returns three objects, precisely, an Apk object, an array of DalvikVMFormat and Analysis object.  Each object has its purpose. The Apk object involves all the contextual information of an application, such as package name and permissions. As for the DalvikVMFormat, it corresponds to the DEX files in an Apk. As for the Analysis object, it provides all the details of classes, methods, and fields found in an Apk. In this context, we used the DEX object, as it provides all the details related to classes, methods, and fields. For every method in an application, we retrieved all its instructions' opcodes. Next, for mapping each opcode to a vector of real numbers, we used Word2Vec~\cite{datatowardsscience} model with skip-gram method. To be specific, we tried to predict each word's context given a specific opcode. As defined in Section~\ref{w2v}, Word2Vec is an implementation technique of word embedding technology. It employs a neural network with one hidden layer to vectorize words. To implement the Word2Vec model, we used TensorFlow library~\cite{abadi2016tensorflow} which provides a rich set of features that allow users to express data computation in an easy fashion. As parameters, we set the embedding size to be 64 and the window length to 1, as we wanted to take only one word in the chain.
\subsubsection{Basic Block Embedding}

After mapping every instruction's opcode into a single vector. We encoded each basic block to its corresponding vector by taking the mean of all its instructions' vectorial representations. In this step, we directly averaged all opcode embedding occurred in a basic block.
Assuming each basic block $BB$ includes n-opcode and each opcode is represented by a $l$-dimensional vector. %We obtained the basic block's embedding vector as follows:

%\begin{equation}
%\tilde{\vec{BB}} = \frac{\sum_{i=1}^{n}{x_i}}{n}
%\end{equation}
%where $x_i$ represents the $l$-dimensional opcode embedding. The algorithm is illustrated as Algorithm~\ref{algo:mean}.

%\begin{algorithm}
%	\DontPrintSemicolon % Some LaTeX compilers require you to use \dontprintsemicolon instead
%	\KwIn{opcode embedding ${v_i : i\in I}$, a set of opcodes ${i \in I}$ of the instruction.}
%	\KwOut{Basic Block embeddings ${v_b : b \in B}$}
%	\For{all basic block b in B}{
%		\For{all opcodes in basic block b}{
%			$b \gets \sum_{i=1}^{n}x_i$
%			
%			$\tilde{\vec{v_b}} = \frac{b}{n}$
%		}
%		\Return{$\tilde{\vec{v_b}}$}
%	}
%	
%	\caption{{Mean Basic Block Embedding}}
%	\label{algo:mean}
%\end{algorithm}

The output of the basic block embedding step is one single vector that represents a node. 
\subsubsection{Program Dependence Graph Construction}

After mapping each vector to its corresponding vectorial representation, we constructed the program dependence graph for every application in our datasets. As mentioned before, every basic block in an application constitutes a node in its program dependence graph. Accordingly, for every basic block in a method, we vectorized it by averaging of all its instructions' vectors. We extract our program dependence graph with the assistance of Androguard. Further, we built edges by bridging every basic block to its related children. To this end, we used NetworkX~\cite{networkx}, a python library for the creation and manipulation of networks. Further, in a network, two nodes should not have the same naming. To be precise, every node in a graph should have a unique name. In the context of machine language, instructions tend to be repetitive, which may lead to disregarding many nodes. Assuming a method has 2 basic blocks with the following opcodes: $B_1$ = $B_2$ = \{\textit{const}, \textit{move}, \textit{goto}\}, the graph will disregard $B_2$, as the encoding of both basic blocks is identical. To avoid this obstacle, we considered two sections in a node's construction, one for the embedding vector, whereas the other for its identifier. Therefore, for every basic block, we selected its parents' details, mainly the corresponding class and method descriptions.   

\subsubsection{Program Dependence Graph Embedding}
%The construction of the program dependence graph results in a set of nodes, each with its corresponding vector. This is not adequately sufficient to capture the structural traits of the entire graph network. For a meaningful data representation, nodes' and edges' structural identities should be integrated to form an embedded network representation that will be used to feed our neural network. In other words, to be able to feed our structural-based neural network, every constructed PDG should be converted to a low dimensional vector space, in which all the graph's properties are fully conserved. 
Consequently, after the program dependence graph's construction, we transformed the entire graph to a vector of real numbers. To this end, we used Structure2Vec~\cite{song2018structure2vec} as a graph embedding framework. Based on the graph topology, Structure2Vec technique aggregates recursively vertex features. Then, the network generates a new embedding of features for every vertex. As shown in Algorithm~\ref{algo:ge}, Structure2Vec based-network starts with an initial embedding at vertex 0 and updates the embedding at every iteration. To obtain the final vector embedding, we took the average of all iterations results. The Structure2Vec process takes two inputs, precisely, an array of nodes' embedding vectors that capture their related properties and adjacency matrix of the program dependence graph, which gives information about the graph's topology, mainly, nodes' connections. In a nutshell, Structure2Vec process is executed as follows:

\begin{algorithm}
	\DontPrintSemicolon % Some LaTeX compilers require you to use 
	\KwIn{Basic block embedding ${v_i : i\in I}$, program dependence graph $g_f$}
	\KwOut{Program dependence graph embedding ${v_p : b \in B}$}
	Initialize $\mu_v^{0} = \vec{Rand}, for all  v \in V$
	
	\For{t=1 to T}{
		\For{$v \in V$}{
			$l_v = \sum_{u \in N(v)} \mu_u^{(t-1)}$
			
			$\mu_v^{(t)} = tanh(W_1x_v + \sigma(l_v))$
		}
	}
	$v_f$ = $W_2(\sum_{v \in V} \mu_v^{T})/len(V))$
	
	\Return{$v_f$}
	\caption{{\sc Program }Dependence Graph Embedding}
	\label{algo:ge}
\end{algorithm}

\subsubsection{Structure-based Neural Network}
\label{neuralnetworkdesign}
After the graph embedding phase, we designed a Multi-Layer Perceptron (MLP), which is a neural network with nodes as neurons with logistic activation. Our aim is to identify whether an application is benign or malicious. Thus, our detection approach is a binary classification process. Accordingly, we designed our neural network to have 64 input neurons, which is equal to the embedding size of an application's program dependence graph. As for the output, we select 2 neurons, one for the benign class, whereas the other for the malware class. Regarding the number of hidden layers, although these layers are not directly interacted with the external environment, we selected 2 hidden layers to tremendously impact the final output. In a nutshell, our neural network can be expressed as the following formula:

\begin{equation}
Y = (X*W_1 + B_1)*W_2 + B_2
\end{equation}

$W_1$ and $W_2$, represent by
$wi_1,wi_2 \in R_p$,
the weights of two hidden layers of our neural network. $bi_1,bi_2 \in R_p$ are the offsets from the origin of the vector space. In this setting, based on the activation function's output,  a program dependence graph $G_h$ can be classified as malicious
if $f(G_h) > 0$  or benign if $f(G_h) < 0$.

%\subsection{Native-code Malware Detection}
\subsection{Native-code based Malware Detection }
Nowadays, several Android malware detectors make use of static analysis techniques. They operate only at the bytecode level, which enables malicious payloads embedded in native code to go invisible as a type of obfuscation technique. As a result, to achieve complete analysis of Android applications, we need a second subsystem that operates on the native code level. In this section, we discuss the structural native-code malware detection subsystem which deals with the native code of Android applications. As illustrated in Figure~\ref{struct}, our second subsystem is a four-step process.

%\begin{figure*}
%	\includegraphics[scale=0.5]{figures/nativecode_diagram}
%	\caption{Inter-language malware detection model}
%	\label{struct_native}
%\end{figure*}

\subsubsection{FCG Extraction}
As a first step, we disassembled Android applications and extracted function call graph based on Angr~\cite{angr}, a python framework for analyzing binaries. 

\subsubsection{Function Embedding}
As a second step, to catch the function call graph topology, we need to consider each function as a node. Normally, a function can involve one or more basic blocks. Each basic block can contain one or more Dalvik instructions. Therefore, our target is to consider every instruction in a function. To this end, we utilized the SIF~\cite{arora2016simple} network to compute the function embedding by using the sequence of instructions vectors generated by the Word2Vec method.
 
\begin{algorithm}
	\DontPrintSemicolon % Some LaTeX compilers require you to use \dontprintsemicolon instead
	\KwIn{Instruction embedding ${v_i : i\in I}$, a set of instructions, parameter $\alpha$ and estimated probabilities ${p(i): i \in I}$ of the instruction.}
	\KwOut{Function embeddings ${v_f : f \in F}$}
	\For{all function f in F}{$
		v_f \gets \frac{1}{|f|} \sum_{i\in f} \frac{\alpha}{\alpha + p(i)}v_i
		$}
	Form a matrix X whose columns are ${v_f : f \in F}$, and let $v$ be its singular vector \\
	\For{all function f in F}{$v_f \gets v_f - uu^Tv_f$}
	\Return{$v_f$}
	\caption{{\sc Simple} Function Embedding}
	\label{algo:sfe}
\end{algorithm}

Based on natural language processing (NLP), given the discourse vector $c_f$, the probability of an instruction which is emitted from the function f is modeled by,
\begin{equation}
\label{model_1}
Pr[i \notin f | \tilde{c_f}] = \alpha p(i) + (1 - \alpha) \frac{exp(<c_f,v_i>)}{Z_{\tilde{c_{f}}}},
\end{equation}
where $\tilde{c_f = \beta c_0 + (1-\beta)c_f},c_0 \perp c_f$, where $\alpha$ and $\beta$ are scalar hyperparameters, and $Z_{\tilde{c_{f}}}$ = $\sum_{i\in f} exp(<\tilde{c_f},v_i>)$ is the normalizing constant. 

The function embedding will be defined as the max likelihood estimate for the vector $c_f$ that generated it. By the model (\ref{model_1}), the likehood for the function is 
\begin{equation}
p[f | c_f] = \prod_{i\in f} p (i|c_f) = \prod_{i\in f} [\alpha p(i) + (1 - \alpha) \frac{exp(<c_f,v_i>)}{Z}].
\end{equation}
Let denote the log likelihood of function f as follows:
\begin{equation}
f_i(\tilde{c_f}) = \log[\alpha p(i) + (1 - \alpha) \frac{exp(<c_f,v_i>)}{Z}].
\end{equation}
After gradient computation and Taylor expansion, we get the maximum likelihood estimator for $\tilde{c_f}$ on the unit sphere is:
\begin{equation}
argmax\sum_{i\in f} f_i(\tilde{c_f})\propto \sum_{i\in f} \frac{\alpha}{p(i) + \alpha}, \alpha = \frac{1-\alpha}{\alpha Z}
\end{equation}
That is the max likelihood estimate is a weighted average of the vectors of the instructions in one function. To estimate $c_f$, we computed the first principal component of $c_f$ for a set of functions as the direction of $c_0$. We present our simple function embedding method in Algorithm~\ref{algo:sfe}.

\subsubsection{Function Call Graph Embedding}
After getting every function's vector, we converted the graph's nodes and edges into vector space to feed our neural network-based classifier. To this end, similarly to program dependence graph embedding, we used Structure2Vec as an embedding framework. Algorithm~\ref{algo:ge} describes our function call graph embedding.
\subsubsection{Native Code Malware detection}
After the graph embedding step, we designed our multi-layer perceptron (MLP). Similarly to structural malware detection, our approach is a binary classification process. Thus, in our work, we followed the JN-SAF ~\cite{Wei:2018:JPE:3243734.3243835} methodology and label all the native libraries included in the following malware families (Boqx, Droidkungfu, Gumen, Lotoor, Ogel, Slembunk, Updtkiller, and Vikinghorde) as malware.
%For benign samples, we selected 633 system libraries from Raspberry Pi B with Raspbian system.
For benign samples, we use all labeled native libraries like the JN-SAF ~\cite{Wei:2018:JPE:3243734.3243835}.
\subsection{Ensemble mechanism}
After getting the outputs of two multi-layer perception(MLP), we have designed an ensemble algorithm to combine the two MLP's results during the training and testing step. The weight of two MLP are trained at the training step. %We set \textit{0} as the \textit{benign} of MLP's output, and \textit{1} as the \textit{malware}.  %Table~\ref{ensembleAlg} presents our OR logic gate algorithm. Only both of two MLP classifier outputs \textit{0}, our malware detection system detects the corresponding sample as \textit{Benign}; otherwise, if one of the two MLP classifiers outputs \textit{1}, our malware detection system detects the corresponding sample as \textit{Malware}. 
%\begin{table}
%	\centering
%	\begin{tabular}{|c|c|c|}
%		\hline
%		MLP-bytecode &MLP-Native & Results  \\ \hline
%		  0 &   0 &     0 \\ \hline
%		 0 &   1 &      1\\ \hline
%		 1 & 0 & 1\\ \hline
%		 1 & 1 & 1\\ \hline
%	\end{tabular}
%	\caption{Ensemble algorithm of two-MLP classifiers.\\ 0 - benign, 1 - malware}
%	\label{ensembleAlg}
%\end{table}

\section{Evaluation}
In this section, we evaluate the performance of our layered Android malware detection approach. We apply and check our proposed model against two public malware datasets, namely DREBIN~\cite{arp2014drebin} and AMD~\cite{wei2017deep}, and one self-collected malware dataset based on VirusTotal~\cite{virustotal} API. As for the benign samples, we used AndroZoo~\cite{allix2016androzoo}.
\subsection{Datasets and Metrics}

\begin{itemize}
	\item \textbf{Drebin:} A well-known dataset with 5560 malicious applications gathered from August 2010 to October 2012. Most Android malware detection research works make use of this dataset~\cite{arp2014drebin}. 
	\item  \textbf{AMD:} It stands for Android Malware Dataset. It contains 24553 malicious applications gathered from 2010 to 2016, consisting 71 malware families. AMD involves highly sophisticated malware applications. It gives clear insights about the recent sophistication techniques applied by malware authors~\cite{wei2017deep}.
	\item  \textbf{AndroZoo-Malware:} Besides aforementioned malware datasets to evaluate the classification issue, we also consider AndroZoo as malware source  in order to solve the biased datasets issues (mentioned by TESSERACT~\cite{pendlebury2019tesseract}), both from the temporal and spatial biases. Following the guideline of TESSERACT, we select malware and 
	spatial biases. Following the guideline of TESSERACT, we select malware and benign samples ratio with 10\% and 90\% (to solve spatial bias), both for the training and testing. In addition, 
	we select our samples from 2017.12.30 to 2019.12.30 (to solve temporal bias). Additionally, we also use VirusTotal to prepare our dataset. But we set our threshold as follows: 
	p~\footnote{AndroZoo's metadata reports the number of p frameworks of positive anti-virus reports on VirusTotal for application} $=$ 0 for benign and p $\geq$ 4 for malware.
	%p~\footnote{AndroZoo's metadata reports the number of p frameworks of positive anti-virus reports on VirusTotal for application} $=$0 for benign and p $≥$ 4 for malware.
	In total, we find out 5,000 malware samples.
	\item  \textbf{AndroZoo-Benign:} A huge collection of Android application gathered from various sources. It involves more than 10 millions applications~\cite{allix2016androzoo}. We used VirusTotal~\cite{virustotal} to scan applications with over 70 antivirus scanners. Additionally, we used a threshold of 4 scanners to label samples. That is, if more than 4 scanners mark an application as malware, we labeled it a malicious. Otherwise, it is considered benign. Our scanning process resulted in 65,000 benign applications.
\end{itemize}
All in all, our dataset can be summarized as Table~\ref{dataset}, which includes 35,110 malware and 65,000 benign samples.
\begin{table}
	\centering
	\begin{tabular}{|c|c|c|}
		\hline
		Malware Samples &Benign Samples & Total  \\ \hline
		35,113 &   65,000 &     100,113 \\ \hline
	\end{tabular}
	\caption{Dataset}
	\label{dataset}
\end{table}
\subsection{Structural-based Classifier Evaluation}
\begin{wrapfigure}{R}{0.4\linewidth}
	\centering
	\vspace{-30pt}
	\includegraphics[width=\linewidth]{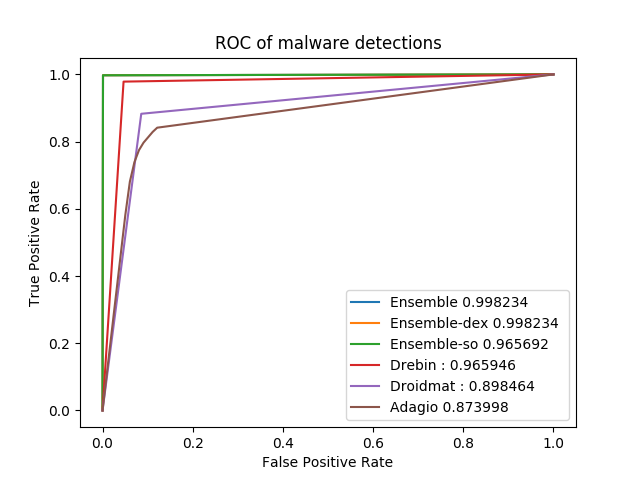}
	\caption{Comparison with other works}
	\label{rocstructure}
	%\vspace{-50pt}
\end{wrapfigure}

In our structural classifier, we evaluated our malware detection system with the following parameters. We set our \textit{batch\_size} to 1. This means, we have trained and tested our classifier with only one program dependence graph in each step. Also, we set our \textit{epochs} to 30 to achieve optimal training. Further, the used datasets were randomly split into 70\% training and 30\% testing. In the same manner, as the first layer, we concentrated on the accuracy, precision, recall, F1. As shown in Table~\ref{Accurancy_graph}, our testing experiments yield an accuracy of more than 99\% on both datasets. This indicates the efficiency of the vectorial representation of the program dependence graph on identifying the structural malicious patterns in an application. Meanwhile, other metrics, such as recall, F1 are also producing significant results.
Figure~\ref{rocstructure} presents the ROC characteristic of our structural-based classifier.

\begin{table*}
	\begin{center}
		\begin{tabular}{|c|c|c|c|c|c|} 
			\hline
			%Benign & 1500 & cell9& &&& \\ 
			%Metric& Drebin &AMD \\ \hline
			%Accuracy & 99.18  &99.43\\ \hline
			%Precision & 99.61 & 99.71 \\ \hline
			%Recall &98.78&99.06\\ \hline
			%F1 &99.20 &99.39\\ \hline
			Metric& Byte-code &Native-code&Ensemble & Drebin&Droidmat\\ \hline
			Accuracy & 99.86  &96.66 & 99.86 &96.58&89.87 \\ \hline
			Precision & 99.95 & 99.06 &99.95 &95.37&90.89 \\ \hline
			Recall &99.82&93.78&99.82 &97.85& 88.28 \\ \hline
			F1 &99.88 &96.35&99.88 &96.59& 89.56 \\ \hline
			
		\end{tabular}
	\end{center}
	
	\caption{Results of structural classifiers (\% representation)}
	\label{Accurancy_graph}
\end{table*}
We can see from the Table~\ref{Accurancy_graph} that the results for the Ensemble is same with Byte-code based method. The reason is that we did not find any benign applications(byte-code) which hides their malicious behaviors into the native code, but the byte-code level act as benign(this result is same with the ~\cite{jayakumargoing}) . At the same time, we also find that most applications which act as malware at native code, have multiple native code libraries, but only one or two of them act malicious behaviors. 
%\begin{figure}
%\hfill
%\subfigure[Drebin]{\includegraphics[scale=0.36]{figures/results_drebin.png}}
%\hfill
%\subfigure[AMD]{
%\centering
%\includegraphics[scale=0.46]{figures/classification_pdg_compare.png}%}
%\hfill
%\caption{Comparison with other works}
%\label{rocstructure}
%\end{figure}
%
%\subsubsection{Native code-based classifier evaluation}

Taking into account that no previous related work has been done on the Drebin dataset, we extracted native libraries from the AMD dataset only. As a result, we obtained 16,387 .so files including seven different architectures namely ARM64, ARM32v7, MIPS, MIPS-r2, X86, X64. Further, all the native libraries that belong to the following families: Boqx, Droidkungfu, Gumen, Lotoor, Ogel, Slembunk, Updtkiller, and Vikinghorde, are considered as malicious native libraries. Moreover, 1,191 of native libraries has ARM architecture. The rest of the architectures has been skipped, as they involve a small number of native libraries. In this experiment, we split our dataset into 70\%training data and 30\% testing data.

As shown in Table~\ref{Accurancy_graph}, Our second classifier achieves an accuracy of 96.6\%. Compared to the structural layer, the performance has dropped. This could be linked to the used dataset's size, as we could not get enough information from previous related works.

%\begin{table}[H]
%	\centering
%	\begin{tabular}{|c|c|c|c|c|}
%		\hline
%		& Precision &Recall  & F1-score&Accuracy\\ \hline
%		SIF-FE&  94.12  &  90.32    & 91.92 & 91.32 \\ \hline
%		
%	\end{tabular}
%	\caption{Results of inter-language classifier (results with \% representation)}
%	\label{result_native_lib}
%\end{table}
To contrast with our method, we get 96.6\% AUC with the Drebin method and 89.85\% with Droidmat on our mixed dataset. Additionally, Adagio gets a lower AUC value, around 87.39\%.\footnote{The results of Drebin, Droidmat, and Adagio are little different from the original work because of the mixed datasets.}
\subsection{Features Processing and Training Overhead}

All work was done on \textit{Euklid}, a SUSE Linux server with 128Gb RAM's size. Table~\ref{featuretime} illustrates the training time of both structural and inter-language classifiers. We can see that our inter-language MLP is trained much faster than our structural MLP. Regarding features processing, taking a look at Table~\ref{featuretime}, it is seen that FCG extraction is nearly three times faster than the PDG's construction. This is related to the number of nodes a graph may involve. A PDG may contain more than 60,000 nodes, which requires a large amount of hardware and time. 

\begin{table}[h]
	\begin{center}
		\begin{tabular}{|c|c|} 
			\hline
			Metric&  \\ \hline
			PDG Construction & 50.49s \\ \hline %& 62.13s \\ \hline
			FCG Construction & 13.03s \\ \hline%& 12.56s \\ \hline
			MLP byte-code structural & 11.68s   \\ \hline
			MLP native-code structural & 09.13s  \\ \hline
			
		\end{tabular}
	\end{center}
	\caption{Average processing time by one application}
	\label{featuretime}
\end{table}
\subsection{Robustness of Structural-Based Approach}
With the fruits of neural networks are unquestionable, recent researches have demonstrated that they fail to exhibit robustness against adversarial inputs. Adversarial samples are usually generated from tweaking the regular inputs through small and calculated perturbations to prompt the model to intentionally output wrong classifications.
\subsubsection{Crafting Adversarial Malware Inputs}
In Android malware detection, the aim of adversarial inputs is to cause the classifier to change its output based on the attackers' goals. Further, our structural-based approach depends on feeding the neural network with an application's program dependence graph embedding. Hence, our primary target is to craft malicious embedding vectors to dupe the model into misclassifying inputs. Generally, when it comes to crafting a malicious graph, there exist several techniques to construct adversarial samples, namely, N-strongest nodes~\cite{xumanis} which is based on selecting and injecting nodes with biggest impacts on the adjacent nodes till the misclassification is achieved. Another commonly used technique is Random Node Injection~\cite{xumanis}, this latter is based on the random injection of nodes into a graph. In our work, to obtain crafted malicious inputs, we used the Gradient-based approach.
\subsubsection{Gradient-Based Approach}

Our adversarial attack focuses on increasing the misclassification rate. As mentioned in the previous sections, we feed the structural-based neural network with the program dependence graph embedding, which is represented as a numeric vector that can be manipulated to obtain the desired output. To this end, our work involves calculating the gradient as well as adding the appropriate perturbations, which results in a 4 steps process:
\begin{enumerate}
\item  We calculated the gradient through the following loss expression:
\begin{equation}
f(X) = ((X*W_1 + B_1)*W_2 + B_2 - Y)^2
\end{equation}

where X is the program dependence graph embedding vector, $W_1$ and $W_2$ are the network weights, $B_1$ and $B_2$ are the network offsets, and $Y$ is the label. In a neural network, the gradient is usually used as an optimization algorithm to minimize the cost with respect to weights for better adjustments. Since our objective is to craft malicious malware samples, we need to construct embedding vectors such that the loss is maximized. In this case, instead of weights, we calculated the gradient by taking the partial derivative of the loss function with respect to X, which is the program dependence graph embedding vector. 
\begin{equation}
\upsilon =  \nabla f(X)/X
\end{equation}
\item We took the gradient sign to know which way to follow to maximize the loss, either by increasing or decreasing the embedding vector. 
\item In this step, we multiplied the gradient sign by epsilon $\epsilon$, which is a small value constant to make sure that our constructed perturbation does not exceed the loss function margin.
\item  Our crafted embedding is obtained from the addition of the original (PDG) embedding vector with perturbation $\eta$.
\end{enumerate}
\begin{wrapfigure}{R}{0.4\linewidth}
	\centering
	\vspace{-30pt}
	\includegraphics[width=\linewidth]{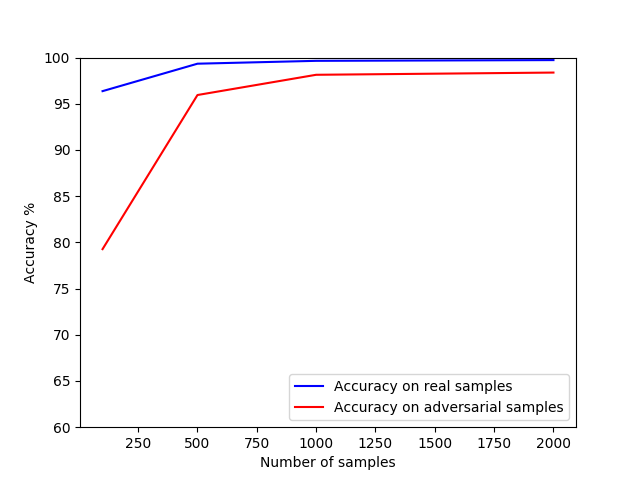}
	\caption{Malware categorization ROC curve of gradient boosting}
	\label{Accuracy of adversarial samples}
	%\vspace{-50pt}
\end{wrapfigure}
%
%\begin{figure}
	%\hfil
	%\centering
	%\subfigure{
	%\includegraphics[scale=0.4]{figures/adv_drebin.png}%}
	%\hfill
	%\subfigure[AMD]{\includegraphics[scale=0.4]{figures/adv_amd.png}}
	%\hfill
	%\caption{Accuracy of adversarial samples}
	%\label{adv}
%\end{figure}  
\subsubsection{Model Robustness Evaluation}

With our neural network has yielded high accuracy on the used datasets, their accuracy drops slightly in the presence of adversarial perturbations to test samples. As shown in Figure~\ref{adv}, as the test set grows in size, the accuracy on the adversarial inputs moves closer to the accuracy achieved on normal test samples. Our experiment involves 4 testing stages, each with different dataset sizes. We selected 100, 500, 1000 and 2000 as size metrics to show the difference in the accuracy's behavior on both normal and adversarial samples. %for both Drebin and AMD datasets.
 With 100 samples, the accuracy drops by nearly 17\% %and 13\%, for both Drebin and AMD, respectively. 
 Next, the accuracy rate increases proportionally to the size of the dataset. At stage 4, with 2000 samples as a test set, for both datasets, the drop rate has reached 1\%, which proves that our neural network is robust against the crafted adversarial samples.

\section{Related Works}
Over the recent years, the detection of Android malware has been under the spotlight of security researchers. Besides the traditional methods such as signature-based and code matching methods, a huge amount of work has been invested into proposing various Android malware detection using machine learning techniques. In the following, we first discuss related works on the structural detection of Android malware using graphs. Next, we proceed to present techniques specifically related to inter-language analysis of Android applications.
\subsection{Android Malware Detection}
Drebin~\cite{arp2014drebin} presents a broad static analysis to detect the Android malware. It evaluates 123,453 applications and 5,560 malware samples. Drebin detects 94\% of the malware with few false alarms and requires 10 seconds for an analysis on average.

In-the-lab and In-the-wild validation~\cite{allix2016empirical} presents several machine learning classifier that only on a set of features build from applications' CFGs. It shows that in the lab the performance outperforms existing machine learing-based approaches. Meanwhile, in the wild scenarios the performance gap, especially F-measure dropping from over 0.9 in the lab to below 0.1.

Mamadroid~\cite{onwuzurike2017mamadroid} presents an Android malware detection system that relies on app behavior. It builds a behavior model, in the form of a Markov chain, from the sequence of abstracted API calls performed by an app, and uses it to extract features and perform classification. By evaluating on a dataset of 8.5 K benign and 35.5K malicious apps, Mamadroid shows that it not only effectively detects malware(with up to 99\% F-malware), but also that the model built by the system keeps its detection capabilities for long periods of time (on average, 86\% and 75\% F-measure, respectively, one and two years after training).

CWLK~\cite{narayanan2016contextual} presents a method to enrich the feature space of a graph kernel that inherently captures structural information with contextual information.It applies the feature-enrichment idea on a state-of-the-art graph kernel, namely, Weisfeiler-Lehman Kernel(WLK) to obtain the Contextual Weisfeiler-Lehman Kernel(CWLK). CWLK evaluates with more than 50,000 Android apps, and outperforms two graph-kernels and three malware detection techniques by more than 5.27\% and 4.87\% F-measure, respectively, while maintaining high efficiency.

CDGDroid~\cite{xu2018cdgdroid} uses the semantics graph representations, that is, control flow graph(CFG), data flow graph(DFG), and their possible combinations, as the features to characterise Android applications. It encodes the graphs in to matrics, and use them to train the classification model via Convolutional Neural Network(CNN). In more detail, given a CFG(or DFG) g, CDGDroid encodes it into a matrix A with size 222*222(total opcodes number in the Android Dalvik-bytecode list) as follows: for each edge (n1,n2) $\in$ g.E, CDGDroid add the element $A[op(n_1)][op(n_2)]$ by 1, where A is initialized as a zero matrix with size 222*222 and op(n) returns the opcode of the node n.
 
\subsection{Structural Detection of Android Malware }
Recently, building an obfuscation-resilient Android malware detection has become an urgent demand. Accordingly, many structural approaches have been proposed to identify malicious patterns via control flow graphs or call graphs. For example, Gascon et al.~\cite{anderson2011graph} present a method for malware detection based on efficient embedding of function call graphs with an explicit feature map inspired by a linear-time graph kernel. They conducted experiments on 12,158 malware samples, achieving an accuracy of 89\%. 

Additionally, Ma et al.~\cite{ma2019combination} present an approach of Android malware classification based on supervised learning on the network properties of control flow graphs. Their evaluation was performed on 10010 benign applications and 10,683 malicious applications, attaining 98.98\% detection precision with high accuracy. In our approach, we differentiated ourselves from previous works by employing the PDG which considers the composition of both functions and data entirely. Also, we utilized the graph embedding technique to transform nodes and edges into vector space while preserving their properties. Moreover, the accuracy score of our structure-based neural network was above 99\% on both datasets.
\subsection{Inter-language Analysis of Android Malware}
With nearly all detection approaches focusing on the Manifest attributes as well as the bytecode level of Android applications, malware authors employ various obfuscations techniques on the native code level to create malware variants. In other words, the native code has become a potential threat vector that can go overlooked by existing detection systems. Consequently, 

Several methods have been proposed to operate on the inter-language side of Android applications. For example, Alam et al.~\cite{alam2017droidnative} present DroidNative, an Android malware detector that operates on both native and bytecode level by using specific control flow patterns to reduce the effect of obfuscations and provides automation. Additionally, Wei et al.~\cite{wei2018jn} present a new approach to conduct inter-language dataflow analysis of Android apps and build an analysis framework, called JN-SAF to compute flow and context-sensitive inter-language points-to information in an efficient way.

\section{Limitations}
Any Android malware detection work is immensely proportional to the quality and quantity of the used dataset. In our work, with the process of collecting benign samples is direct, gathering sophisticated Android malware is a demanding task that entails technical efforts. Apart from that, the main limitation of our layered approach is the training phase of our structural layer which employs graph embedding. As we use program dependence graph, the majority of the extracted graphs have a huge amount of nodes which demand a large-size main memory (RAM). Unfortunately, with a RAM of 128 Gb, the process of extracting a graph with more than 90,000 nodes gets killed. Consequently, we disregarded any graph with more than 90,000 total of nodes. Moreover, in terms of robustness evaluation, we had no adequate time to largely assess the robustness of the inter-language layer using adversarial attacks.

\section{Conclusion and Future work}

In this paper, we presented an approach that employs the principle of detection in depth. To be specific, our method blends two detection techniques that handle different aspects of an Android application. Our structure-based layer focuses on the structural side of an Android application, specifically the code's properties and their relationships, whereas the inter-language layer acts on the native code level. Further, the use of graph embedding technique is shown to be efficient, as it led us to achieve high performance. As far as we know, this is a first work to combine both byte-code (DEX file) and native-code (SO file) in Android malware detection. Hence, adapting this technique to other areas, such as desktop malware and phishing detection seems to be an interesting avenue for future research. 
\small{
\bibliographystyle{unsrt}
\bibliography{paper}

\begin{thebibliography}{10}

\bibitem{smartphonemarketshare1}
TEAM COUNTERPOINT.
\newblock Global smartphone market share: By quarter, 2021.

\bibitem{mcafee}
Mcafee labs threats report - june 2021.
\newblock {\em McAfee Inc., Santa Clara, CA. Available:
  https://www.mcafee.com/enterprise/en-us/lp/threats-reports/jun-2021.html},
  2021.

\bibitem{chen2018tinydroid}
Tieming Chen, Qingyu Mao, Yimin Yang, Mingqi Lv, and Jianming Zhu.
\newblock Tinydroid: A lightweight and efficient model for android malware
  detection and classification.
\newblock {\em Mobile Information Systems}, 2018, 2018.

\bibitem{jang2016detecting}
Jae-wook Jang, Jaesung Yun, Aziz Mohaisen, Jiyoung Woo, and Huy~Kang Kim.
\newblock Detecting and classifying method based on similarity matching of
  android malware behavior with profile.
\newblock {\em SpringerPlus}, 5(1):273, 2016.

\bibitem{wei2014amandroid}
Fengguo Wei, Sankardas Roy, and Xinming Ou.
\newblock Amandroid: A precise and general inter-component data flow analysis
  framework for security vetting of android apps.
\newblock In {\em Proceedings of the 2014 ACM SIGSAC Conference on Computer and
  Communications Security}, pages 1329--1341, 2014.

\bibitem{arzt2014flowdroid}
Steven Arzt, Siegfried Rasthofer, Christian Fritz, Eric Bodden, Alexandre
  Bartel, Jacques Klein, Yves Le~Traon, Damien Octeau, and Patrick McDaniel.
\newblock Flowdroid: Precise context, flow, field, object-sensitive and
  lifecycle-aware taint analysis for android apps.
\newblock {\em Acm Sigplan Notices}, 49(6):259--269, 2014.

\bibitem{gordon2015information}
Michael~I Gordon, Deokhwan Kim, Jeff~H Perkins, Limei Gilham, Nguyen Nguyen,
  and Martin~C Rinard.
\newblock Information flow analysis of android applications in droidsafe.
\newblock In {\em NDSS}, volume~15, page 110, 2015.

\bibitem{li2015iccta}
Li~Li, Alexandre Bartel, Tegawend{\'e}~F Bissyand{\'e}, Jacques Klein, Yves
  Le~Traon, Steven Arzt, Siegfried Rasthofer, Eric Bodden, Damien Octeau, and
  Patrick McDaniel.
\newblock Iccta: Detecting inter-component privacy leaks in android apps.
\newblock In {\em 2015 IEEE/ACM 37th IEEE International Conference on Software
  Engineering}, volume~1, pages 280--291. IEEE, 2015.

\bibitem{lu2012chex}
Long Lu, Zhichun Li, Zhenyu Wu, Wenke Lee, and Guofei Jiang.
\newblock Chex: statically vetting android apps for component hijacking
  vulnerabilities.
\newblock In {\em Proceedings of the 2012 ACM conference on Computer and
  communications security}, pages 229--240, 2012.

\bibitem{anderson2011graph}
Blake Anderson, Daniel Quist, Joshua Neil, Curtis Storlie, and Terran Lane.
\newblock Graph-based malware detection using dynamic analysis.
\newblock {\em Journal in computer Virology}, 7(4):247--258, 2011.

\bibitem{xu2018cdgdroid}
Zhiwu Xu, Kerong Ren, Shengchao Qin, and Florin Craciun.
\newblock Cdgdroid: Android malware detection based on deep learning using cfg
  and dfg.
\newblock In {\em International Conference on Formal Engineering Methods},
  pages 177--193. Springer, 2018.

\bibitem{narayanan2016contextual}
Annamalai Narayanan, Guozhu Meng, Liu Yang, Jinliang Liu, and Lihui Chen.
\newblock Contextual weisfeiler-lehman graph kernel for malware detection.
\newblock In {\em 2016 International Joint Conference on Neural Networks
  (IJCNN)}, pages 4701--4708. IEEE, 2016.

\bibitem{arp2014drebin}
Daniel Arp, Michael Spreitzenbarth, Malte Hubner, Hugo Gascon, Konrad Rieck,
  and CERT Siemens.
\newblock Drebin: Effective and explainable detection of android malware in
  your pocket.
\newblock In {\em Ndss}, volume~14, pages 23--26, 2014.

\bibitem{xu2021detecting}
Peng Xu, Claudia Eckert, and Apostolis Zarras.
\newblock Detecting and categorizing android malware with graph neural
  networks.
\newblock In {\em Proceedings of the 36th Annual ACM Symposium on Applied
  Computing}, pages 409--412, 2021.

\bibitem{xufalcon}
Peng Xu, Claudia Eckert, and Apostolis Zarras.
\newblock Falcon: Malware detection and categorization with network traffic
  images.

\bibitem{xu2021hawkeye}
Peng Xu, Youyi Zhang, Claudia Eckert, and Apostolis Zarras.
\newblock {HawkEye: Cross-Platform Malware Detection With Representation
  Learning on Graphs}.
\newblock In {\em International Conference on Artificial Neural Networks
  (ICANN)}, 2021.

\bibitem{hybroid}
Mohammad~Reza Norouzian, Peng Xu, Claudia Eckert, and Apostolis Zarras.
\newblock Hybroid: Toward android malware detection and categorization with
  program code and network traffic.
\newblock pages 259--278. Springer International Publishing, 2021.

\bibitem{suarez2017droidsieve}
Guillermo Suarez-Tangil, Santanu~Kumar Dash, Mansour Ahmadi, Johannes Kinder,
  Giorgio Giacinto, and Lorenzo Cavallaro.
\newblock Droidsieve: Fast and accurate classification of obfuscated android
  malware.
\newblock In {\em Proceedings of the Seventh ACM on Conference on Data and
  Application Security and Privacy}, pages 309--320, 2017.

\bibitem{dong2018understanding}
Shuaike Dong, Menghao Li, Wenrui Diao, Xiangyu Liu, Jian Liu, Zhou Li, Fenghao
  Xu, Kai Chen, XiaoFeng Wang, and Kehuan Zhang.
\newblock Understanding android obfuscation techniques: A large-scale
  investigation in the wild.
\newblock In {\em International Conference on Security and Privacy in
  Communication Systems}, pages 172--192. Springer, 2018.

\bibitem{hu2009large}
Xin Hu, Tzi-cker Chiueh, and Kang~G Shin.
\newblock Large-scale malware indexing using function-call graphs.
\newblock In {\em Proceedings of the 16th ACM conference on Computer and
  communications security}, pages 611--620. ACM, 2009.

\bibitem{sun2014nativeguard}
Mengtao Sun and Gang Tan.
\newblock Nativeguard: Protecting android applications from third-party native
  libraries.
\newblock In {\em Proceedings of the 2014 ACM conference on Security and
  privacy in wireless \& mobile networks}, pages 165--176. ACM, 2014.

\bibitem{dalvikfig}
Jayesh Akojwar.
\newblock Dalvik virtual machine, 2013.

\bibitem{ferrante1987program}
Jeanne Ferrante, Karl~J Ottenstein, and Joe~D Warren.
\newblock The program dependence graph and its use in optimization.
\newblock {\em ACM Transactions on Programming Languages and Systems (TOPLAS)},
  9(3):319--349, 1987.

\bibitem{kazmierski2011system}
Tom~J Ka{\'z}mierski and Adam Morawiec.
\newblock {\em System Specification and Design Languages: Selected
  Contributions from FDL 2010}, volume 106.
\newblock Springer Science \& Business Media, 2011.

\bibitem{datatowardsscience}
Dhruvil Karani.
\newblock Introduction to word embedding and word2vec, 2018.

\bibitem{mikolov2013distributed}
Tomas Mikolov, Ilya Sutskever, Kai Chen, Greg~S Corrado, and Jeff Dean.
\newblock Distributed representations of words and phrases and their
  compositionality.
\newblock In {\em Advances in neural information processing systems}, pages
  3111--3119, 2013.

\bibitem{datatowardssciencesentence}
Oliver Borchers.
\newblock Sentence embeddings. fast, please!, 2019.

\bibitem{datatowardssciencegraph}
Primož Godec.
\newblock Graph embeddings — the summary, 2018.

\bibitem{perozzi2014deepwalk}
Bryan Perozzi, Rami Al-Rfou, and Steven Skiena.
\newblock Deepwalk: Online learning of social representations.
\newblock In {\em Proceedings of the 20th ACM SIGKDD international conference
  on Knowledge discovery and data mining}, pages 701--710. ACM, 2014.

\bibitem{grover2016node2vec}
Aditya Grover and Jure Leskovec.
\newblock node2vec: Scalable feature learning for networks.
\newblock In {\em Proceedings of the 22nd ACM SIGKDD international conference
  on Knowledge discovery and data mining}, pages 855--864. ACM, 2016.

\bibitem{kipf2016semi}
Thomas~N Kipf and Max Welling.
\newblock Semi-supervised classification with graph convolutional networks.
\newblock {\em arXiv preprint arXiv:1609.02907}, 2016.

\bibitem{song2018structure2vec}
Le~Song.
\newblock Structure2vec: Deep learning for security analytics over graphs.
\newblock 2018.

\bibitem{narayanan2017graph2vec}
Annamalai Narayanan, Mahinthan Chandramohan, Rajasekar Venkatesan, Lihui Chen,
  Yang Liu, and Shantanu Jaiswal.
\newblock graph2vec: Learning distributed representations of graphs.
\newblock {\em arXiv preprint arXiv:1707.05005}, 2017.

\bibitem{detectionindepth}
Gavin Reid.
\newblock Detection in depth, 2016.

\bibitem{basicblock}
JavaPoint.
\newblock Basic block, 12 2019.

\bibitem{abadi2016tensorflow}
Mart{\'\i}n Abadi, Paul Barham, Jianmin Chen, Zhifeng Chen, Andy Davis, Jeffrey
  Dean, Matthieu Devin, Sanjay Ghemawat, Geoffrey Irving, Michael Isard, et~al.
\newblock Tensorflow: A system for large-scale machine learning.
\newblock In {\em 12th $\{$USENIX$\}$ Symposium on Operating Systems Design and
  Implementation ($\{$OSDI$\}$ 16)}, pages 265--283, 2016.

\bibitem{networkx}
NetworkX.
\newblock Software for complex networks, 12 2019.

\bibitem{angr}
angr.
\newblock Angr, 2 2020.

\bibitem{arora2016simple}
Sanjeev Arora, Yingyu Liang, and Tengyu Ma.
\newblock A simple but tough-to-beat baseline for sentence embeddings.
\newblock 2016.

\bibitem{Wei:2018:JPE:3243734.3243835}
Fengguo Wei, Xingwei Lin, Xinming Ou, Ting Chen, and Xiaosong Zhang.
\newblock Jn-saf: Precise and efficient ndk/jni-aware inter-language static
  analysis framework for security vetting of android applications with native
  code.
\newblock In {\em Proceedings of the 2018 ACM SIGSAC Conference on Computer and
  Communications Security}, CCS '18, pages 1137--1150, New York, NY, USA, 2018.
  ACM.

\bibitem{wei2017deep}
Fengguo Wei, Yuping Li, Sankardas Roy, Xinming Ou, and Wu~Zhou.
\newblock Deep ground truth analysis of current android malware.
\newblock In {\em International Conference on Detection of Intrusions and
  Malware, and Vulnerability Assessment (DIMVA'17)}, pages 252--276, Bonn,
  Germany, 2017. Springer.

\bibitem{virustotal}
Gaurav Sood.
\newblock {\em virustotal: R Client for the virustotal API}, 2017.
\newblock R package version 0.2.1.

\bibitem{allix2016androzoo}
Kevin Allix, Tegawend{\'e}~F Bissyand{\'e}, Jacques Klein, and Yves Le~Traon.
\newblock Androzoo: Collecting millions of android apps for the research
  community.
\newblock In {\em 2016 IEEE/ACM 13th Working Conference on Mining Software
  Repositories (MSR)}, pages 468--471. IEEE, 2016.

\bibitem{pendlebury2019tesseract}
Feargus Pendlebury, Fabio Pierazzi, Roberto Jordaney, Johannes Kinder, and
  Lorenzo Cavallaro.
\newblock $\{$TESSERACT$\}$: Eliminating experimental bias in malware
  classification across space and time.
\newblock In {\em 28th $\{$USENIX$\}$ Security Symposium ($\{$USENIX$\}$
  Security 19)}, pages 729--746, 2019.

\bibitem{jayakumargoing}
Sudeep~Nanjappa Jayakumar.
\newblock Going native: Using a large-scale analysis of android apps to create
  a practical native-code sandboxing policy.

\bibitem{xumanis}
Peng Xu, Bojan Kolosnjaji, Claudia Eckert, and Apostolis Zarras.
\newblock Manis: Evading malware detection system on graph structure.
\newblock In {\em The 35th ACM/SIGAPP Symposium on Applied
  Computing(SAC‘20)}, pages 1688--1695.

\bibitem{allix2016empirical}
Kevin Allix, Tegawend{\'e}~F Bissyand{\'e}, Quentin J{\'e}rome, Jacques Klein,
  Yves Le~Traon, et~al.
\newblock Empirical assessment of machine learning-based malware detectors for
  android.
\newblock {\em Empirical Software Engineering}, 21(1):183--211, 2016.

\bibitem{onwuzurike2017mamadroid}
Lucky Onwuzurike, Enrico Mariconti, Panagiotis Andriotis, Emiliano
  De~Cristofaro, Gordon Ross, and Gianluca Stringhini.
\newblock {MaMaDroid: Detecting Android Malware by Building Markov Chains of
  Behavioral Models (Extended Version)}.
\newblock {\em ACM Transactions on Privacy and Security (TOPS)}, 22(2):14,
  2019.

\bibitem{ma2019combination}
Zhuo Ma, Haoran Ge, Yang Liu, Meng Zhao, and Jianfeng Ma.
\newblock A combination method for android malware detection based on control
  flow graphs and machine learning algorithms.
\newblock {\em IEEE Access}, 7:21235--21245, 2019.

\bibitem{alam2017droidnative}
Shahid Alam, Zhengyang Qu, Ryan Riley, Yan Chen, and Vaibhav Rastogi.
\newblock Droidnative: Automating and optimizing detection of android native
  code malware variants.
\newblock {\em computers \& security}, 65:230--246, 2017.

\bibitem{wei2018jn}
Fengguo Wei, Xingwei Lin, Xinming Ou, Ting Chen, and Xiaosong Zhang.
\newblock Jn-saf: Precise and efficient ndk/jni-aware inter-language static
  analysis framework for security vetting of android applications with native
  code.
\newblock In {\em Proceedings of the 2018 ACM SIGSAC Conference on Computer and
  Communications Security}, pages 1137--1150. ACM, 2018.

\end{thebibliography}
}
\end{document}